# Corpus non alignés et ADT. Essai de comparaison entre les présidents français et brésiliens de l'ère contemporaine

Carlos Maciel[1], Damon Mayaffre[2], Laurent Vanni[3]

[1] UCA, CNRS, BCL – Carlos.Maciel@univ-cotedazur.fr

[2] CNRS, UCA, BCL – Damon.Mayaffre@univ-cotedazur.fr

[3] CNRS, UCA, BCL – Laurent.Vanni@univ-cotedazur.fr

## Abstract

Is there an ADT method that can deal with non-aligned bilingual corpora? Does the textual genre exert a sufficiently strong constraint on the discourse that would make texts written in different languages comparable, provided they are of identical genre? To answer these two questions, one methodological, the other linguistic, this contribution gathers in a single corpus French and Brazilian presidential speeches of the contemporary era (1950-2020), from de Gaulle to Macron, from Kubitschek to Lula, i.e. 15 million words. A methodological path is proposed from the simple frequency dictionary to the factorial treatment of the cooccurrencial profiles of words, in order to establish a generic transnational presidential speech.

## Résumé

Existe-t-il une méthode ADT susceptible de traiter des corpus bilingues non alignés ? Le genre textuel exerce-t-il une contrainte suffisamment forte sur le discours qui rendrait comparable des textes écrits dans des langues différentes sous condition d'être de genre identique ? Pour répondre à ces deux questions, une méthodologique, l'autre linguistique, cette contribution rassemble dans un même corpus les discours politiques présidentiels français et brésiliens de l'époque contemporaine (1950-2020), de de Gaulle à Macron, de Kubitschek à Lula, soit 15 millions de mots. Un parcours méthodologique est proposé du simple dictionnaire des fréquences jusqu'au traitement factoriel des profils concurrentiels des mots, afin d'établir un parler présidentiel générique transnational.

**Mots clés :** discours présidentiel, logométrie, genre textuel, isotopie, corrélats, cooccurrences.

## 1. Introduction

L'ADT [Iezzi et al. 2020, Lebart et Salem 1994 ; Lebart, Pincemin et Poudat 2019] analyse depuis plusieurs décennies principalement des *corpus contrastifs monolingues*. Le corpus de textes, partitionné en sous-corpus selon une certaine métadonnée (l'auteur, la date, le genre…), constitue alors la référence ou la norme statistico-linguistique, par rapport à laquelle les sous-corpus en question peuvent se singulariser (calcul de spécificités, calcul de distances intertextuelles, etc.). Différemment, [Zimina 2000], aux JADTs de Lausanne, puis dans diverses publications, a proposé une méthodologie ADT, implémentée dans le logiciel *Lexico*, pour étudier statistiquement des *corpus bilingues alignés*, répandus dans le TAL [Véronique *éd*. 2000] : il s'agissait alors de comparer grâce aux outils textométriques des textes jumeaux et parallélisés – de strictes traductions – écrits dans des langues différentes, par exemple les textes officiels canadiens rédigés en français et en anglais.

Soit, ici, aujourd'hui, ni un corpus contrastif monolingue, ni un corpus bilingue aligné, mais un corpus composé de textes que l'analyste juge comparables, et pourtant écrits dans des langues différentes, par des locuteurs différents et dans des situations distinctes : *un corpus bilingue non parallélisable*. Que peut en dire l'ADT en complément de la thèse HDR de [Kraif 2014] ?

Les discours des présidents de la République français et des présidents de la République brésiliens ont été ainsi rassemblés sur une longue période historique 1950-2020 dans un corpus français/portugais non aligné de 15 millions de mots[1].

L'hypothèse linguistique forte que nous mettons à l'épreuve est empruntée à [Brunet 2009], dans la lignée historique des travaux de Volochinov ou Bakhtine : le genre discursif – ici le genre « discours présidentiel » – n'est pas seulement un pallier intermédiaire entre la langue et le discours, mais un pallier transversal aux langues, qui relie les discours. Autrement dit, le genre contraint ou informe les productions langagières, par-delà les langues, et rend comparable la production discursive d'un Macron ou d'un Kubitschek avec celle d'un de Gaulle ou d'un Lula[2].

A titre introductif, nous concédons que (1) la comparabilité problématique que nous éprouvons, au titre d'un genre commun, grâce à l'approche quantitative, n'apparait envisageable que par la proximité étymologique ou culturelle des deux langues envisagées (le français et le portugais brésilien) ; que (2) les méthodes ADT que nous proposons peinent à être strictement formalisées : il s'agit d'avantage d'un parcours méthodologique « bricolé » au sens de Levy Strauss, convoquant l'AFC, le traitement de la cooccurrence ou le simple dictionnaire de fréquences que d'une méthode intégrée.

## 2. La contrainte du genre : un stock lexical commun

Si les calculs aussi éprouvés sur corpus contrastif monolingue que le *z-score* ou les *spécificités* apparaissent impuissants pour traiter nos données bilingues non alignées, un simple détour par les dictionnaires de fréquence respectifs des sous-corpus français et brésilien apparait instructif des contraintes génériques qui pèsent sur les discours.

Après étiquetage et lemmatisation[3], le logiciel *Hyperbase* sélectionne hiérarchiquement les 20 substantifs les plus utilisés, respectivement par les présidents français et les présidents brésiliens. Or 18 sur 20 apparaissent communs aux deux ensembles, parfois au rang près d'utilisation (figure 1).

Ainsi, par exemple, nos présidents placent au premier rang 'pays'/'país', et au deuxième rang 'année'/'ano'. Plus loin dans le tableau, 'chose', 'problèmes', 'monde', etc. en français, correspondent strictement en portugais à 'coisa', 'problema', 'mundo' etc.

Il existe donc un stock lexical fortement contraint du discours présidentiel que l'on pourrait qualifier d'épilexique générique, indépendant de la langue et du pays. Et de part et d'autre de l'Atlantique, lorsqu'on est président de la République, on parle à ses concitoyens de la 'vie'/'vida', du 'travail'/'trabalho', du 'temps/'tempo', de 'président'/'presidente', etc.

---

[1] Le sous-corpus français rassemble les grands textes de de Gaulle, Pompidou, Giscard, Mitterrand, Chirac, Sarkozy, Hollande et Macron, entre 1958 et 2021, équivalant 4,5 millions de mots. Le sous-corpus brésilien, plus important en taille, rassemble la plupart des discours de Kubitschek, Goulart, Castelo Branco, Médici, Geisel, Figueiredo, Sarney, Collor, Franco, Cardoso, Lula, entre 1956 et 2010, équivalant 11 millions de mots.

[2] Précisons encore que le Brésil et la France ont des régimes présidentiel et semi-présidentiel semblables qui donnent à la parole présidentielle française et brésilienne une autorité identique.

[3] Le corpus a été lemmatisé et étiqueté morphosyntaxiquement par les logiciels Spacy et TreeTagger dont on sait les limites de performance mais qui ont l'avantage de fonctionner sur le français et le portugais.

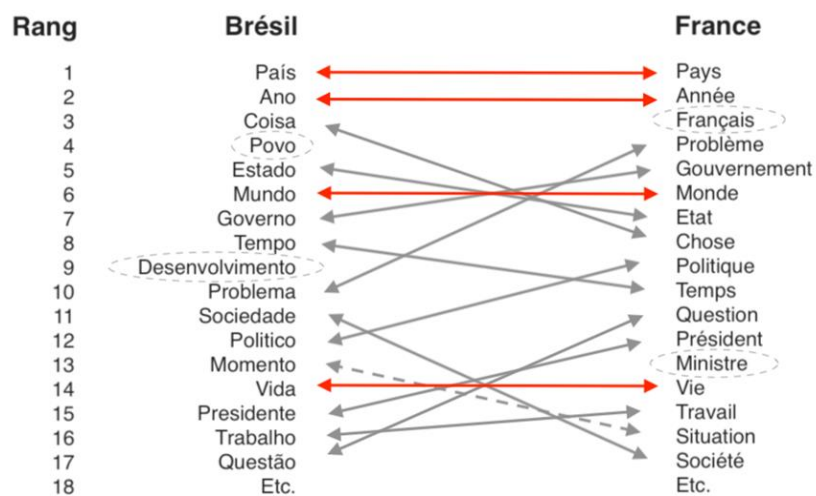

*Figure 1. Etude comparée du dictionnaire de fréquences (Brésil/France). Substantifs classés par leur rang de fréquences*

Dans cette stéréotypie lexicale des discours présidentiels que révèle le tableau des fréquences, il deviendrait important de commenter les très rares exceptions, puisqu'elles soulignent des particularismes nationaux France/Brésil suffisamment forts pour résister au genre présidentiel. Par exemple, les acteurs principaux des discours en France sont les 'Français' (3ème rang), là où le discours brésilien ne convoque pas les 'Brésiliens' et semble s'adresser avant tout au peuple ('povo', 4ème rang). Il y aurait ainsi un patriotisme mieux ancré dans les habitudes lexicales au sein du vieux cadre national français, là où la jeune République fédérale se sert d'un vocabulaire plus politique que patriotique. Par ailleurs, le terme de 'desenvolvimento' très haut placé au Brésil (9ème rang) n'a pas d'équivalent dans les 20 premiers substantifs en France ('développement') et marque sans doute, sur la longue durée, la problématique d'un pays en pleine expansion politique, démographique, économique, sociale durant la période, là où la France post-30 glorieuses et post-coloniale se vit peut-être davantage sur le déclin ou la préservation plutôt que dans le développement

## 3. La contrainte du politique : des thématiques croisées

Des mots aux thèmes, l'ADT propose depuis [Benzécri 1973] dans son principe fondamental puis [Viprey 1997] dans une application circonstanciée, l'étude des thématiques du discours – appelées *isotropies* et implémentées dans le logiciel *Hyperbase* sous le nom de *Corrélats* : seraient-elles identiques ou voisines dans les sous-corpus français et brésiliens ?

Pour chaque pays, nous construisons une matrice de contingence carré (ou triangulaire puisque symétrique à la diagonale) composée en ligne des 300 substantifs les plus utilisés du discours (dont les 20 mentionnés plus haut) et en colonne ces mêmes substantifs ; dans les cellules du tableau, le nombre de fois où les mots considérés cooccurrent. Le traitement factoriel des profils coocurrentiels des mots ainsi constitués donne les deux représentations suivantes. S'il nous apparait hasardeux mathématiquement de prétendre à une comparaison formelle des deux sorties machines[4], l'exploration visuelle des deux nuages permet de conclure à des ressemblances évidentes, sans doute imputables au genre commun.

---

[4] Par exemple, le fait que la contribution aux axes 1 et 2 se chiffrent respectivement à 6,1% et 3,2% dans de corpus français et à 8,5 et 6,3% dans le corpus brésilien nous parait difficilement interprétable sans prendre de gros risques sur-interprétatifs.

Le nuage de mots français se lit facilement en 4 isotropies majeures selon des résultats que nous avons souvent décrits et analysés [Mayaffre 2012]. Le discours présidentiel français a ainsi une isotropie ou composante (1) institutionnelle ('majorité', 'assemblée', 'mandat', etc.), une composante (2) économique ('déficit', 'prix', 'banque', 'agriculture', etc.), une composante (3) sociale et sociétale ('chômage', 'retraite', 'contrat' mais aussi 'enfant', 'éducation', 'famille') et une composante (4) internationale ('arme', 'paix', 'guerre', 'conflit', 'puissance', etc.).

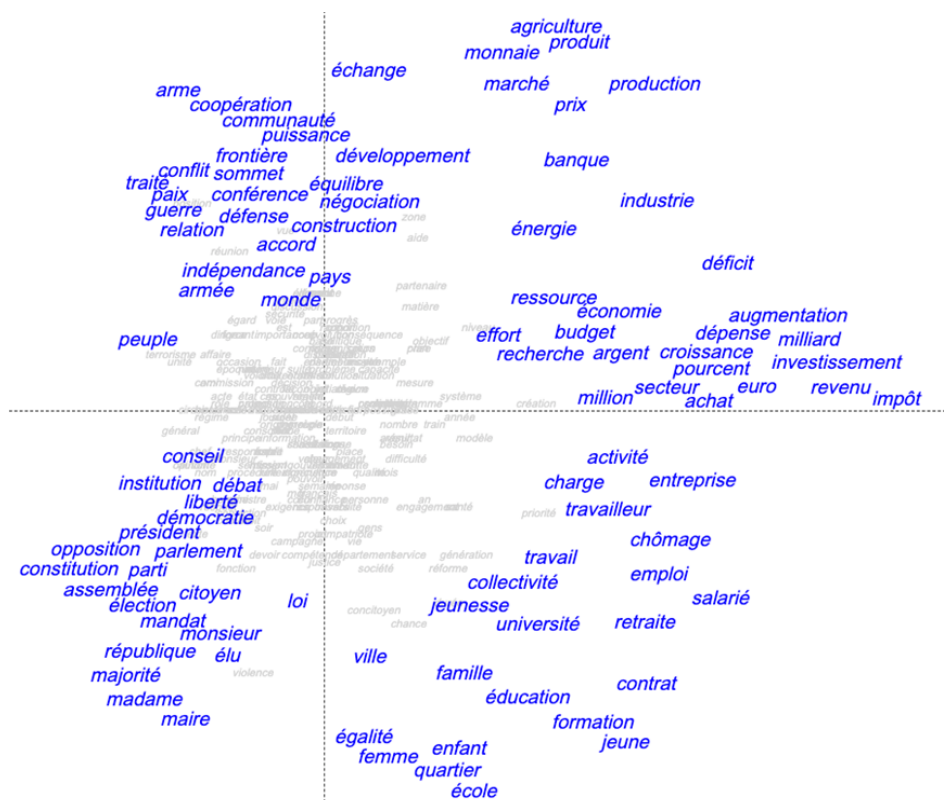

*Figure 2. Corrélats des 300 substantifs le plus utilisés dans le corpus français (Hyperbase – 2022)*

Dans le discours brésilien, nous semblons retrouver l'essentiel de cette information thématique – même si nous soulignerons les nuances. Nous repérons ainsi facilement l'isotropie (1) institutionnelle ('ministro', 'governador', 'deputado' 'prefeito'), l'isotropie (2) économique ('dólar', 'taxa', 'inflação'), et l'isotropie (3) sociale et sociétale avec 'emprego', 'salário', 'pobre', 'família', 'escola' ou 'universidade' etc. Les mots peuvent différer entre les deux pays, mais la structuration thématique ou isotropique des textes du corpus apparait identique.

Cependant l'isotropie de politique internationale très prégnante en France ne semble pas organiser le discours présidentiel brésilien, comme il structure le discours Français. La quatrième composante brésilienne compte ainsi des termes moins géopolitiques et plus politiques qu'en France (comme nous l'avions pressenti avec 'povo' vs. 'Français'). Ainsi, par exemple, au 'conflit' en France semble répondre l''integração' au Brésil, à la 'puissance' française semble répondre le 'desenvolvimento' brésilien. Là où en France nous enregistrons 'guerre' ou 'arme', etc., nous enregistrons au Brésil 'liberdade', 'democracia', etc. Cette différence ne semble pas devoir surprendre au regard de l'histoire et de la position comparées du Brésil et de la France dans le concert international. La France fait partie des puissances nucléaires, le Brésil non. La France fait partie de manière permanente du Conseil de sécurité, le Brésil non. La France est un ancien pays colonial qui garde une forte influence en Afrique, le Brésil non. Du reste, d'autres facteurs explicatifs pourraient être avancés parmi lesquels les prérogatives du président accordées par les Constitutions respectives (chef des armées en

France comme l'écrit hautement l'article 15 de la Constitution, moins revendiqué au Brésil) et par la culture politique différente (le Brésil ne fait pas de la guerre un point central de son action internationale, là où l'affirmation de la puissance française sur les différents continents est une constante des politiques françaises aux XX$^{ème}$ et XXI$^{ème}$ siècle.).

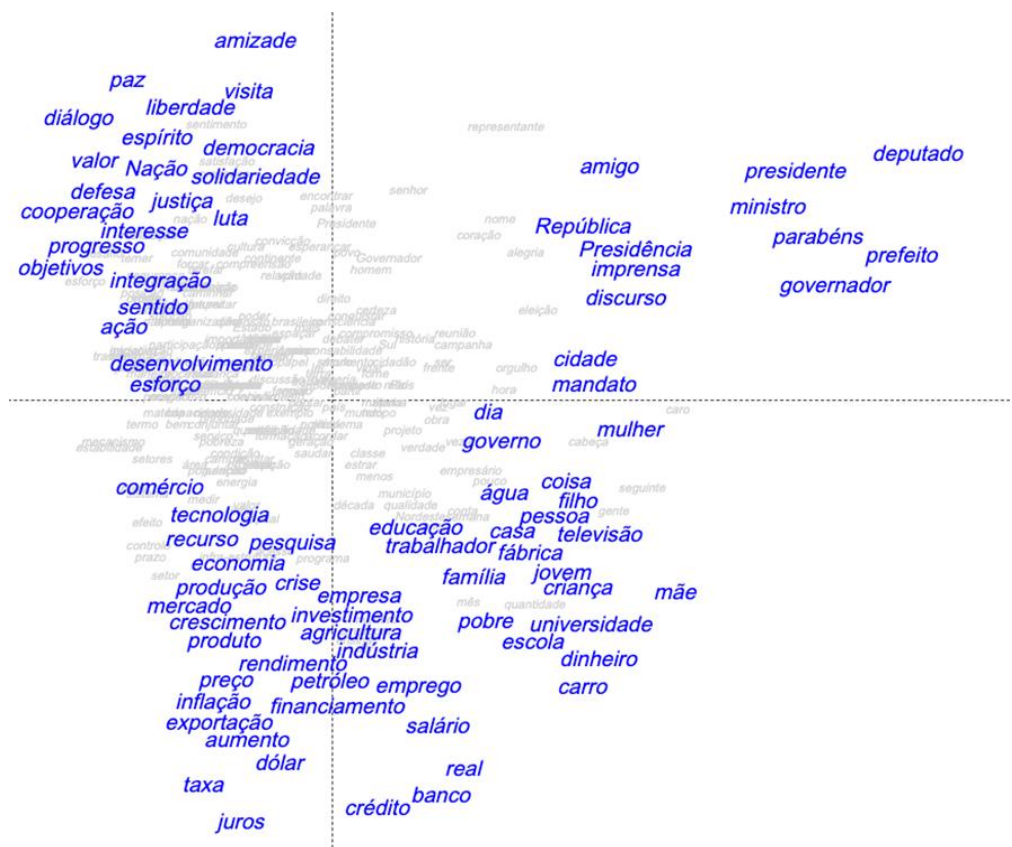

*Figure 3. Corrélats des 300 substantifs les plus utilisés dans le corpus brésilien (Hyperbase 2022)*

## 4. Des thèmes au sens : les limites d'une comparaison multilingue

Des mots aux thèmes, des thèmes au sens, nous proposons pour conclure de zoomer sur l'utilisation respective d'un terme partagé et très utilisé dans le deux sous-corpus : 'travail' / 'trabalho'. Nous rappelons avec la linguistique contextuelle anglo-saxonne [par ex. Halliday et Hasan 1976] et française [par ex. Guiraud 1960] que le sens d'un mot est la somme de ses contextes d'utilisation. Et précisons avec [Mayaffre 2014] que cette affirmation signifie, statistiquement, que le sens d'un mot peut être considéré comme *la somme de ses cooccurrences*.

Partant, le nuage des cooccurrents du mot 'travail'/'trabalho' laisse paraitre des nuances sémantiques fortes entre le France et le Brésil (figure 4). Certes on remarquera quelques mots cooccurrents identiques 'fruit'/'fruto', 'travailleur'/'trabalhador', 'emploi'/'emprego' ou d'une certaine manière 'entreprise'/'oficina'.

Néanmoins, le sens de travail apparait très incarné au Brésil. Il s'adresse aux familles ('família'), aux enfants ('criança'), aux hommes et aux femmes ('homem' et 'mulher') ou encore à la main d'œuvre ('mao-de-obra') ou aux équipes ('equipe'). Le travail au Brésil se définit ainsi avant tout comme le produit d'une activité humaine. En France les acteurs du travail semblent avoir disparu du discours (à l'exception de 'travailleur'), et la question du travail se technicise autour de revendications sociales historiques ('cotisation', 'branche',

'salaire', 'réforme', etc.). En France, le travail est avant tout regardé comme une organisation économique et sociale. De manière plus obvie, le travail relève au Brésil, pays où le secteur agro-industriel tient une place dominante dans l'économie, du travail des champs ('campo', 'cana', 'canavial', 'sisal') et soulève la question essentielle du partage des terres ('terra').

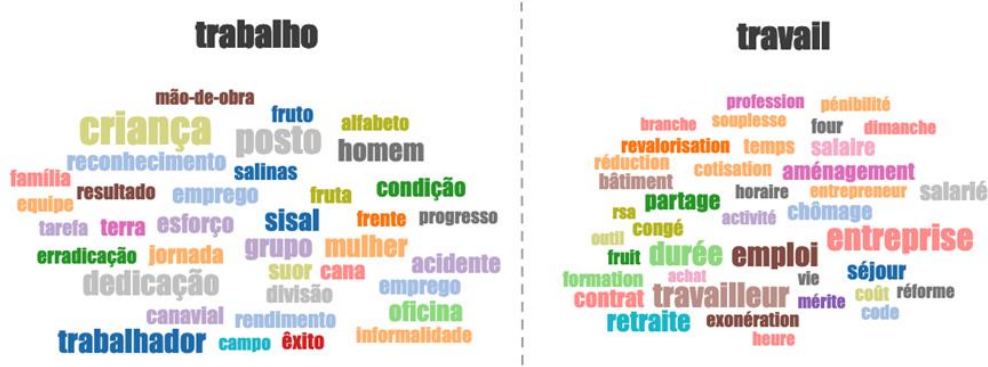

*Figure 4. Cooccurrents de 'travail'/'trabolho' (la taille des mots est proportionnelle à l'indice de cooccurrence)*

## 4. Conclusion

La comparaison de corpus bilingues non alignés ne semble pas pouvoir être automatisée. Le passage par une langue-pivot est inenvisageable puisqu'il abolirait les nuances sémantiques que le linguiste se propose précisément d'étudier. Dès lors, les outils ADT peuvent être convoqués sur chaque sous-corpus : charge alors à l'analyste de mettre en comparaison des sorties-machines différentes mais de même facture (profils cooccurrentiels, rang de fréquences, AFC, etc.). Encore faut-il que les sous-corpus comparés soient comparables. A défaut d'une langue commune et de locuteurs identiques, cette contribution a montré que le genre offrait cette comparabilité. Le genre transcende la langue et les auteurs. Il unifie ici notre corpus et rend analogues les prises de parole des présidents français ou brésiliens qui parlent au nom de leur pays. L'expérience ici réalisée sur le lexique pourrait être plus marquée encore sur d'autres observables linguistiques, sensibles aux genres, d'ordres grammaticaux ou syntaxiques.

## Références

Zimina M. (2000). "Alignement de textes bilingues par classification ascendante hiérarchique" in *JADT 2000*, Lausanne, 171-178.